\documentclass[prl,a4paper,twocolumn,superscriptaddress]{revtex4}

\usepackage{amsmath}
\usepackage{amssymb}
\usepackage{textcomp}
\usepackage{upgreek}
\usepackage{units}
\usepackage[pdftex]{graphicx}
\usepackage{multirow}
\usepackage{microtype}
\usepackage{enumitem}
\usepackage{xcolor}

\usepackage{txfonts}

\renewcommand{\vec}[1]{\ensuremath{\boldsymbol{#1}}}

\begin{document}

\title{Magnon nodal-line semimetals and drumhead surface states in anisotropic pyrochlore ferromagnets}

\author{Alexander Mook}
\affiliation{Max-Planck-Institut f\"ur Mikrostrukturphysik, D-06120 Halle (Saale), Germany}

\author{J\"urgen Henk}
\affiliation{Institut f\"ur Physik, Martin-Luther-Universit\"at Halle-Wittenberg, D-06099 Halle (Saale), Germany}

\author{Ingrid Mertig}
\affiliation{Max-Planck-Institut f\"ur Mikrostrukturphysik, D-06120 Halle (Saale), Germany}
\affiliation{Institut f\"ur Physik, Martin-Luther-Universit\"at Halle-Wittenberg, D-06099 Halle (Saale), Germany}

\begin{abstract}
We introduce a new type of topological magnon matter: the magnonic pendant to electronic nodal-line semimetals. Magnon spectra of anisotropic pyrochlore ferromagnets feature twofold degeneracies of magnon bands along a closed loop in reciprocal space. These magnon nodal lines are topologically protected by the coexistence of inversion and time-reversal symmetry; they require the absence of spin-orbit interaction (no Dzyaloshinskii-Moriya interaction). We calculate the topological invariants of the nodal lines and show that details of the  associated magnon drumhead surface states depend strongly on the termination of the surface. Magnon nodal-line semimetals complete the family of topological magnons in three-dimensional ferromagnetic materials.
\end{abstract}


\date{\today}

\maketitle

\paragraph{Introduction.} 
Over the recent years, nontrivial topologies of magnon spectra have become a thriving field of research. In striking analogy to \emph{electronic} topological matter \cite{Hasan10}, topological \emph{magnon} matter has been identified. The `drosophilae' of such topological magnon insulators (TMIs) \cite{Zhang2013}, which are the pendant to electronic Chern insulators, are (twodimensional) ferromagnets on a kagome lattice with Dzyaloshinskii-Moriya interaction (DMI) \cite{Dzyaloshinsky58,Moriya60,Katsura2010,Mook14a,Mook14b,Mook15a,Mook15b,Lee2015}. The latter causes complex hopping matrix elements in the free-boson Hamiltonian of magnons and, thus, breaks time-reversal symmetry; this points towards the textured magnetic flux in the Haldane model \cite{Haldane1988}. As a result, Berry curvatures and Chern numbers are nonzero and cause topologically protected edge magnons. The latter revolve unidirectionally the sample in accordance with the bulk-boundary correspondence \cite{Hatsugai1993,Hatsugai1993a}. Recently, Cu-(1,3-benzenedicarboxylate) was identified as a TMI which is very well approximated by the kagome model \cite{Chisnell2015}; TMIs on the honeycomb lattice have been proposed as well \cite{Owerre2016a}.

The quest for topologically nontrivial systems has been initiated by the discovery of the magnon Hall effect \cite{Onose2010,Ideue12} in ferromagnetic pyrochlore oxides, mostly because the transverse thermal Hall conductivity has been related to the Berry curvature of the bulk magnons \cite{Katsura2010,Matsumoto2011,Matsumoto2011a}. The quite natural extension of the topological classification to three-dimensional systems lead to the discovery of magnon Weyl semimetals \cite{Li2016,Mook2016}, in which the crossing points of two magnon bands act as source and sink of Berry flux (again, in close analogy to electronic systems \cite{Wan2011,Xu2015}).

In this Letter, we complete the family of topological magnonic objects in three-dimensional ferromagnetic materials by predicting `magnon nodal-line semimetals' (magnon NLSMs), the magnon pendant of electronic NLSMs \cite{Mikitik2006,Burkov2011a,Yu2015,Weng2015,Kim2015,Huang2016}. For this purpose, we consider a ferromagnetic pyrochlore lattice with anisotropic exchange interactions but without spin-orbit interaction (SOI). We find two nodal lines, that is, two closed loops in reciprocal space along which two magnon bands are degenerate. On top of this, we identify the protecting symmetries and calculate topological invariants of the nodal lines. Magnon spectra for the $(111)$ surface feature drumhead surface states, i.\,e., the hallmarks of NLSMs. The dispersion relations of the latter depend strongly on the termination of the surface. Eventually, we discuss the effect of a nonzero DMI on the spectra and suggest experiments to identify magnon NLSMs.

\begin{figure}
	\centering
	\includegraphics[width=1\columnwidth]{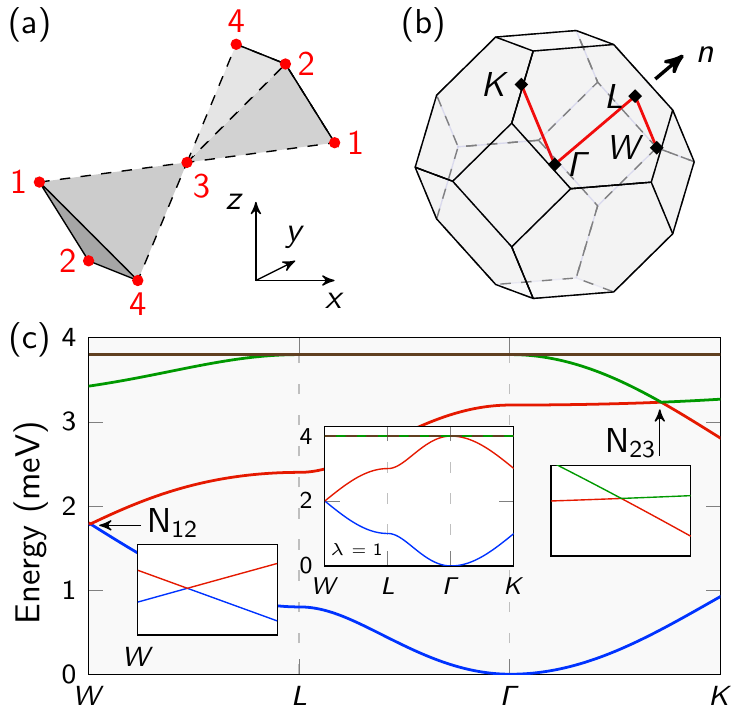}
	\caption{(Color online) (a) Corner-sharing tetrahedra of the pyrochlore lattice with basis sites (red dots, labeled $1, \ldots, 4$). Sites $1$, $2$, and $4$ form a kagome lattice with nearest-neighbor interaction $J_\mathrm{N}$ (solid bonds); site $3$ connects adjacent kagome layers by a weaker interaction $J'_\mathrm{N}$ (dashed bonds). (b) Brillouin zone of the fcc lattice with indicated high-symmetry points and $\vec{n} = (1,1,1) / \sqrt{3}$. (c) Magnon spectrum [$\varepsilon_n (\vec{k})$, $n=1,\ldots,4$] of the anisotropic pyrochlore ferromagnet along the red path depicted in (b) for $J_\mathrm{N} = \unit[1]{meV}$, $\lambda = 0.8$, and $s=1/2$. Band crossings are marked N$_{12}$ and N$_{23}$, with insets showing magnifications. The central inset displays the spectrum for the isotropic case ($\lambda = 1$).}
	\label{fig:lattice}
\end{figure}

\paragraph{Model and spin-wave analysis.} 
The pyrochlore lattice consists of four interpenetrating face-centered cubic (fcc) lattices, resulting in a regular array of corner-sharing tetrahedra [see Fig.~\ref{fig:lattice}(a)]. Along the $[111]$ direction, kagome layers alternate with triangular layers. 

Considering only nearest-neighbor Heisenberg exchange interactions, there is only coupling between sites within a kagome layer [$J_{\mathrm{N}}$, solid bonds in Fig.~\ref{fig:lattice}(a)] and between sites in a triangular layer with sites in the adjacent kagome layers [$J'_{\mathrm{N}}$, dashed bonds in Fig.~\ref{fig:lattice}(a)]. To introduce anisotropic exchange, we assume that $0 < J'_\mathrm{N} < J_\mathrm{N}$. Thus, the Hamiltonian 
\begin{align}
  H = &- J_\mathrm{N} \sum_{\langle ij \rangle}^{\text{kagome}} \vec{s}_i \cdot \vec{s}_j - J'_\mathrm{N} \sum_{\langle ij \rangle}^{\text{triangular}} \vec{s}_i \cdot \vec{s}_j 
  \label{eq:Hamiltonian}
\end{align}
includes only symmetric exchange between spins $\vec{s}_i$ and $\vec{s}_j$ at sites $i$ and $j$, respectively. An external magnetic field is not considered because it merely shifts the magnon spectrum towards higher energies. Moriya's symmetry rules \cite{Moriya60} would in principle allow for a nonzero DMI \cite{Elhajal2005,Kotov2005} but for the time being we consider vanishing SOI\@.

A truncated Holstein-Primakoff transformation \cite{Holstein1940},
\begin{align*}
	s_i^z &\rightarrow s - n_i, \\
	s_i^+ = s_i^x + \mathrm{i} s_i^y &\rightarrow \sqrt{2s}  a_i, \\
	s_i^- = s_i^x - \mathrm{i} s_i^y &\rightarrow \sqrt{2s} a_i^\dagger,
\end{align*}
is applied  (linear spin-wave approximation, i.\,e., no magnon-magnon interactions). The annihilation operators $a_i$ and the creation operators $a_i^\dagger$ obey the boson commutation rules; $n_i = a_{i}^{\dagger} a_{i}$ is the magnon number operator. This low-temperature approximation yields the  free-magnon Hamiltonian matrix 
\begin{align*}
	H_\mathrm{free} = - 2 s J_\mathrm{N} \left( 
		\begin{matrix}
			-2 - \lambda	&	c(y,-z)		& 	\lambda c(x,y)	& 	c(z,-x)			\\
			c(y,-z)	&	-2 - \lambda			&	\lambda c(x,z)	& 	c(x,-y)		\\
			\lambda c(x,y)	& \lambda c(x,z)	&	-3 \lambda	& \lambda c(y,z) 	\\
			c(z,-x)	&	c(x,-y)		&  \lambda c(y,z)	& -2 - \lambda
		\end{matrix}
	\right),
\end{align*}
in which $c(\alpha,\beta) \equiv \cos(\operatorname{sgn}(\alpha) \, k_\alpha + \operatorname{sgn}(\beta) \, k_\beta )$ for $\alpha, \beta = x, y, z$ and $\lambda \equiv J'_\mathrm{N} /  J_\mathrm{N} \le 1$. A non-unitary Bogoliubov transformation is not necessary because of the ferromagnetic ground state. Diagonalization of $H_\mathrm{free}$ 
gives the magnon dispersions $\varepsilon_n(\vec{k})$ and the magnon eigenstates $|u_n(\vec{k})\rangle$. In the following, we present results for $s = 1/2$, $J_\mathrm{N}=\unit[1]{meV}$, and $\lambda = 0.8$; the lattice constant is set to unity.

\begin{figure}
	\centering
	\includegraphics[width=1\columnwidth]{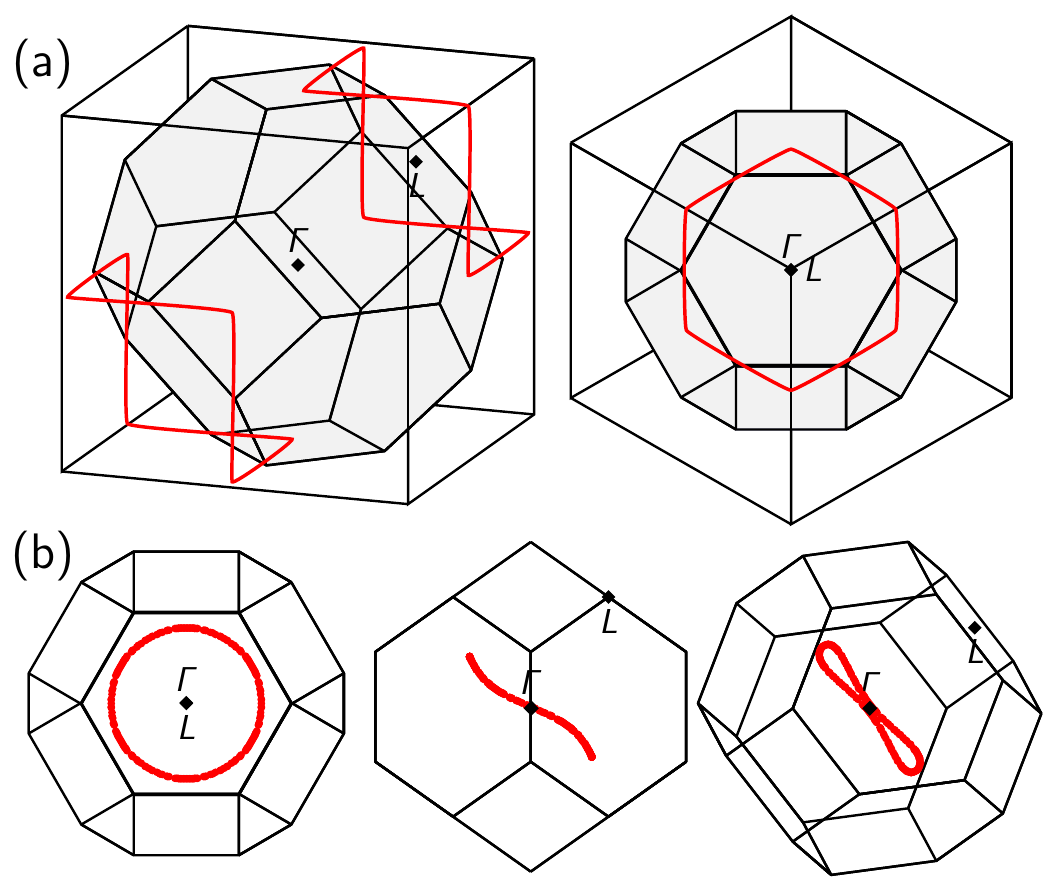}
	\caption{(Color online) Perspective views of the nodal lines (red lines) N$_{12}$ (a) and N$_{23}$ (b) in the fcc BZ\@. The high-symmetry points $\Gamma$ and $\mathrm{L}$ are indicated for all perspectives. Parameters as in Fig.~\ref{fig:lattice}(c).}
	\label{fig:nodal}
\end{figure}

The magnon band structure is made up of four bands, consistent with the four basis atoms of the pyrochlore lattice [Fig.~\ref{fig:lattice}(b) shows the high-symmetry lines along which the spectrum is shown in (c)]. For isotropic exchange ($\lambda = 1$; see central inset), the two topmost bands are dispersionless and the lower two bands are mirror images of each other. For $\lambda < 1$ [$\lambda = 0.8$ in Fig.~\ref{fig:lattice}(c)], one of the formerly dispersionless bands (green) becomes dispersive; this leads to a crossing with band~2 (red; counted from below) at a single $\vec{k}$ point on the $\Gamma$--$\mathrm{K}$ line (marked N$_{23}$). Additionally, the crossing N$_{12}$ of band~1 (blue) and band~2 (red), which is at the $W$ point for $\lambda = 1$, is shifted along the $W$--$L$ line for $\lambda < 1$. These degeneracies are part of nodal lines shown in Fig.~\ref{fig:nodal}.  In the following we label the nodal lines by N$_{\mu \nu}$, in which $\mu$ ($\nu$) stands for the energetically lower (higher) band forming the nodal line.

N$_{12}$ is best depicted on the cube circumscribing the fcc Brillouin zone [BZ; Fig.~\ref{fig:nodal}(a)]. It does not lie exactly on the cube's faces but is set-off by a tiny amount toward the BZ center. For decreasing $\lambda$, N$_{12}$ keeps contracting until it becomes more and more ring-like; it is moved toward the hexagonal faces of the BZ until it vanishes in the $\mathrm{L}$ point for $\lambda = 0$ (which is the limit of noninteracting kagome layers).

N$_{23}$ has its center at the $\Gamma$ point [Fig.~\ref{fig:nodal}(b)], its diameter increases for decreasing $\lambda$. It, too, is not a planar ring in $k$ space but shows modulations that are consistent with the threefold rotational symmetry of the $[111]$ direction ($\Gamma$--$\mathrm{L}$ line).

\paragraph{Symmetry and topology analysis.}
Nodal lines are differentiated by their protecting symmetry \cite{Hirayama2016}. A type-1 nodal line is protected by mirror symmetry and, thus, has to lie within a mirror plane. A type-2 nodal line is protected by the simultaneous presence of inversion and time-reversal symmetry in a system without SOI; it may appear in  generic positions in $k$ space.

\begin{figure}
	\centering
	\includegraphics[width=1\columnwidth]{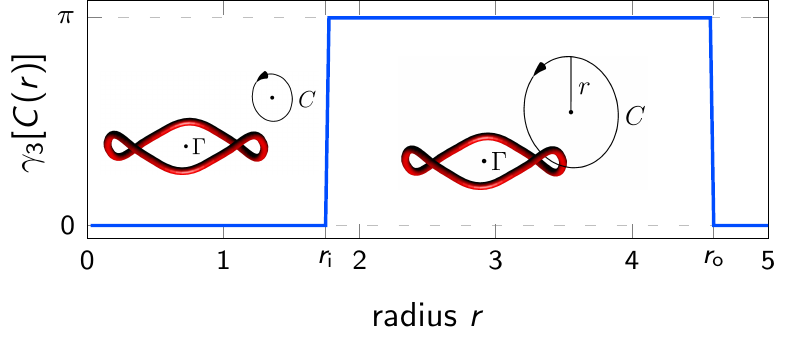}
	\caption{(Color online) Integrated Berry phase of band~3 in dependence on the radius $r$ of the loop $C$. The center of $C$ is chosen well outside the nodal line (left inset). The nodal line (red) becomes `woven' with $C$ if its radius exceeds $r_\mathrm{i}$ (right inset). For $r > r_\mathrm{o}$ the nodal line and $C$ are again separated.}
	\label{fig:berry}
\end{figure}

The Hamiltonian considered here is time-reversal invariant because there is no complex hopping (in contrast to the case with DMI \cite{Katsura2010,Mook2016}); due to the ferromagnetic ground state, inversion symmetry is also present. In combination, these symmetries imply that the Berry curvature \cite{Berry1984,Zak1989}
\begin{align*}
 \Omega_n(\vec{k}) & = \mathrm{i} \langle \nabla_{\vec{k}} u_n(\vec{k})| \times | \nabla_{\vec{k}}u_n(\vec{k}) \rangle
\end{align*}
and, consequently, the Chern numbers
\begin{align*}
 C_n & = \int_S \Omega_n(\vec{k}) \cdot \vec{e} \,\mathrm{d}S
\end{align*}
vanish, thereby suggesting a topologically trivial situation ($\vec{e}$ is the $\vec{k}$-dependent local normal of a closed surface $S$). However, the nontrivial topology of a nodal line is identified by the Berry phase integrated along an arbitrary closed loop $C$ \cite{Huang2016},
\begin{align*}
	\gamma_n[C] & = \mathrm{i} \int_{C} \langle u_n(\vec{k}) | \nabla_{\vec{k}} u_n(\vec{k}) \rangle \,\mathrm{d} \vec{k}.
\end{align*}
If $C$ and the nodal line are intertwined (sketched in Fig.~\ref{fig:berry}),  $\gamma_n[C] = \pi$ (nontrivial), otherwise $\gamma_n[C] = 0$ (trivial).

To prove the nontrivial topology of the nodal lines, we computed $\gamma_n[C(r)]$ of the first and third band in dependence on the radius $r$ of a circle $C$. Figure~\ref{fig:berry} shows as an example $\gamma_{3}[C(r)]$ but we note that the  argument is valid also for $\gamma_1[C(r)]$. The center of $C$ is chosen such that it is well separated from the nodal line N$_{23}$. Hence, the nodal line does not `puncture' the surface enclosed by $C$ and $\gamma_{3}[C(r)] = 0$ (left inset in Fig.~\ref{fig:berry}). Increasing of $r$ leads to the `interweaving' of the nodal line and $C$ (right inset) once the critical radius $r_\mathrm{i}$ is reached: $\gamma_{3}[C(r)] = \pi$. Upon a further increase of $r$ nodal line and loop become separated again and $\gamma_{3}[C(r)]$ falls back to zero for $r > r_\mathrm{o}$.

The above topology analysis tells that the nodal lines are of type~2. We now turn to the drumhead surface states (DSSs) which are another characteristic of nodal lines.

\paragraph{Surface states.} The surface magnon dispersion is analyzed in terms of the spectral density $N_{p}(\varepsilon, \vec{k})$ which is computed for a semi-infinite geometry by Green function renormalization \cite{Henk1993}. The renormalization proceeds as follows. For the chosen (111) surface, the pyrochlore lattice is decomposed into principal layers (PLs) which are parallel to that surface. The principal layers are chosen in such a way that the Hamiltonian matrix of the semi-infinite system comprises only interactions within a PL and among adjacent PLs. In the infinite set of equations for the PL-resolved Green-function matrix the inter-PL interactions are iteratively reduced (renormalized). After a few iterations the entire Hamilton matrix becomes effectively block diagonal which allows to compute the spectral densities
\begin{align*}
	N_{p}(\varepsilon, \vec{k}) & = - \frac{1}{\pi} \lim_{\eta \rightarrow 0^+} 
	\mathrm{Im} \left[ \operatorname{Tr} G_{pp} (\varepsilon + \mathrm{i} \eta, \vec{k}) \right]
\end{align*}
of PL $p$ from the Green function block $G_{pp}$. A finite $\eta$ (here: $\unit[0.001]{meV}$) ensures convergence and introduces broadening. Hence, we have access to the bulk spectral density ($p = \infty$) and to that of any other PL, in particular that of the surface ($p = 0$).

The perspective views given in Fig.~\ref{fig:nodal} in which $\Gamma$ and $\mathrm{L}$ coincide [right in (a); left in (b)] reflect the rotational symmetry of both the nodal lines and the hexagonal BZ of the $(111)$ surface. The argumentation given below is also valid for other surfaces with nontrivial projections of the nodal lines.

\begin{figure}
	\centering
	\includegraphics[width=1\columnwidth]{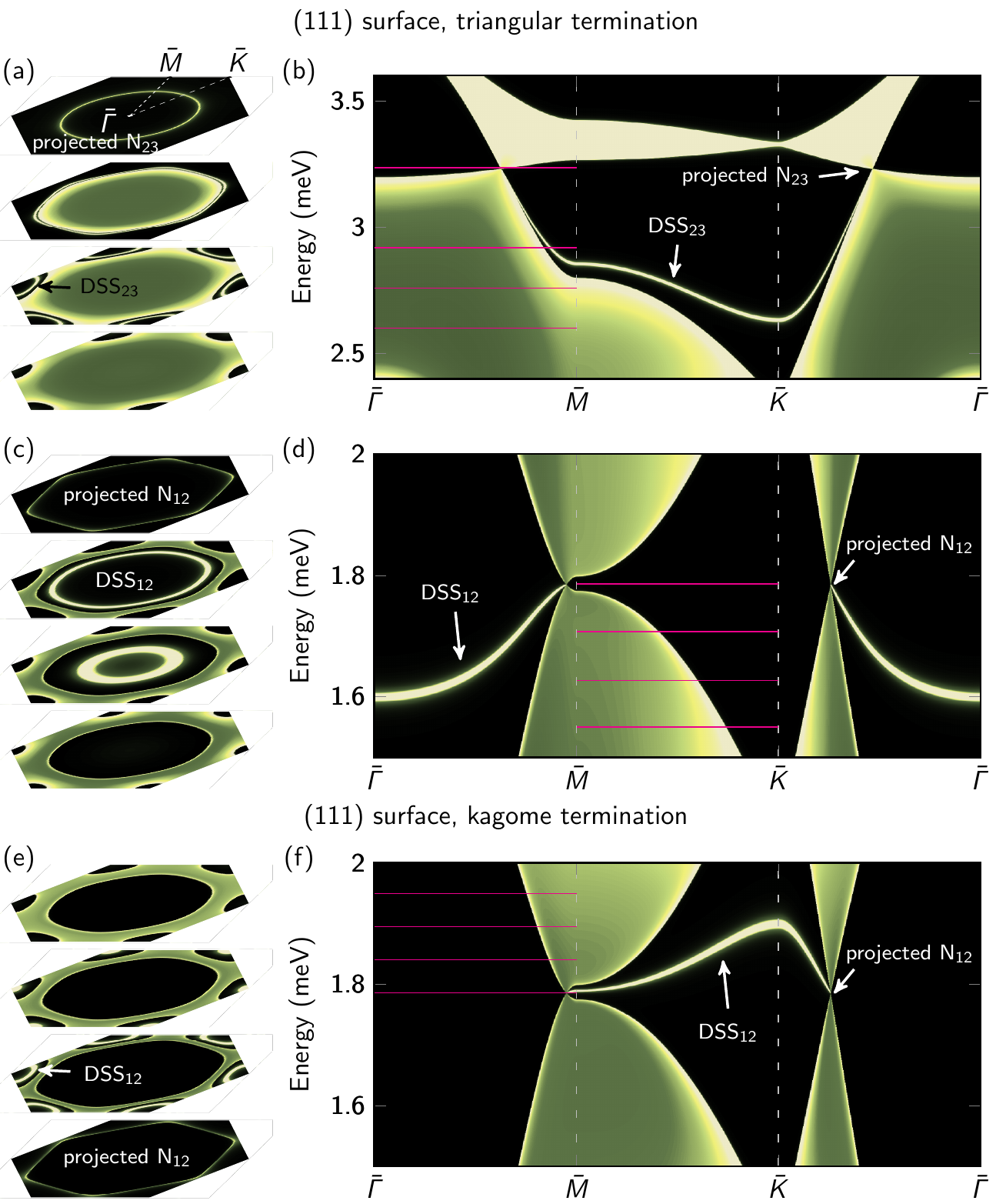}
	\caption{(Color online) Magnons at the $(111)$ surface of an anisotropic pyrochlore ferromagnet for two surface terminations; Panels (a) -- (d) show results for a triangular termination (top and central row), while (e) and (f) show results for a kagome termination (bottom row). The surface spectral density $N_{0}(\varepsilon, \vec{k})$ is represented as color scale (black: zero; white: maximum). Bulk magnons appear as broad features, surface states as sharp light lines. Panels (a), (c), and (e) show constant-energy cuts through the entire surface Brillouin zone for energies indicated by magenta lines in (b), (d), and (f), respectively. Panels (b), (d), and (f) display spectral densities along high-symmetry directions of the surface Brillouin zone. The projected nodal lines N$_{12}$ and N$_{23}$ and the respective drumhead surface states DSS$_{12}$ and DSS$_{23}$ are marked. Parameters as in Fig.~\ref{fig:lattice}(c).}
	\label{fig:surface}
\end{figure}

The bulk magnon spectrum appears as broad features upon projection onto the $(111)$ surface (cf.\ the extended green regions in Fig.~\ref{fig:surface}); surface states show up as comparatively sharp features. Constant-energy cuts of the surface magnon spectrum at energies close to the nodal line N$_{23}$ are plotted in Fig.~\ref{fig:surface}(a) [the corresponding energies are indicated in Fig.~\ref{fig:surface}(b)]. For the lowest energy, there is no surface state and only bulk states are visible [bottom constant-energy cut in (a)]. At the energy for which only N$_{23}$ is present, the projected bulk states form a closed ring in the surface BZ [top constant-energy cut in (a)]. Additionally, DSS$_{23}$ which is associated with N$_{23}$ produces a ring-like feature whose extension `shrinks' toward $\overline{\mathrm{K}}$ with decreasing energy. The cut of the surface magnon spectrum along high-symmetry lines of the surface BZ [Fig.~\ref{fig:surface}(b)] shows two points of N$_{23}$: one on the $\overline{\Gamma}$--$\overline{\mathrm{M}}$ and another on the $\overline{\Gamma}$--$\overline{\mathrm{K}}$ line. DSS$_{23}$ is `suspended' at these points, its considerable dispersion indicates that the `membrane is stretched quite loosely'. Figs.~\ref{fig:surface}(c) and (d) show the same scenario for the projection of N$_{12}$ and DSS$_{12}$: with decreasing energy, DSS$_{12}$ `shrinks' toward $\overline{\mathrm{\Gamma}}$.

The $(111)$ surface of the pyrochlore lattice allows for two terminations: a kagome or a triangular layer. The results discussed up to here were obtained for the triangular termination. Considering the kagome termination [Figs.~\ref{fig:surface}(e) and (f)], there is no difference in the bulk contributions but a major variation in the DSSs, here exemplified by DSS$_{12}$. If we call the region about the $\overline{\mathrm{\Gamma}}$ point the inside of the projected N$_{12}$, we find DSS$_{12}$ now in its exterior. This fundamental dependence on the surface termination has also been observed for electronic NLSM, for example, in the alkaline-earth stannides, germanides, and silicides \cite{Huang2016}. 

\paragraph{Effect of the Dzyaloshinskii-Moriya interaction.}
For the electronic NLSMs, SOI has to be absent \cite{Huang2016,Hirayama2016} or at least very weak because otherwise the nodal lines would be lifted and other topological states could occur. The same is valid for the magnon nodal lines: the pyrochlore lattice allows for a nonzero DMI which would break time-reversal symmetry \cite{Mook2016}. Thus, the inevitable combination of inversion and time-reversal symmetry would be removed and nodal lines could not exist~\footnote{In a recent study, Su et al.\ considered anisotropy in combination with DMI and its effects on magnon Weyl points \cite{Su2016}. However, the authors did not discuss the case of zero DMI and did not identify nodal lines.}. Instead, the magnon bands would carry nonzero Chern numbers and topological surface states would appear. Other topological states are likely, e.\,g., Weyl points \cite{Mook2016,Su2016}.

\paragraph{Experimental considerations.} To prove the existence of magnon nodal lines, one could utilize either a direct or an indirect approach. 

Considering a direct mapping, e.\,g., by inelastic neutron scattering for the bulk magnons \cite{Mena2014} and electron energy loss spectroscopy for the surface magnons \cite{Zakeri2013}, we propose to tune the anisotropy (imbalance of $J_{\mathrm{N}}$ and $J'_{\mathrm{N}}$) by applying strain. As the diameter of the nodal line is sensitive to the ratio $J'_\mathrm{N} / J_\mathrm{N}$, it will in- or decrease accordingly; such a behavior can be detected in experiments. If samples with different surface termination can be grown, one could utilize the strong dependence of the DSSs on the termination [cf.\ Fig.~\ref{fig:surface}(d) and (f)].

In view of an indirect approach, one could think of transverse transport, for instance the magnon Hall effect. However, these  effects require a nonzero Berry curvature which is ruled out by the simultaneous presence of inversion and time-reversal symmetry required for an NLSM\@. We recall that most of the ferromagnetic pyrochlores exhibit a sizable DMI; examples are Lu$_2$V$_2$O$_7$, In$_2$Mn$_2$O$_7$, and Ho$_2$V$_2$O$_7$ \cite{Onose2010,Ideue12}. Thus, the measurement of the magnon Hall effect can be used for the identification of samples with negligible DMI, which are likely to exhibit magnon nodal lines. The Hamiltonian~\eqref{eq:Hamiltonian} is surprisingly simple as it involves only exchange interaction, which suggests that magnon NLSMs could be quite common and not restricted to the pyrochlore lattice. The examination of this conjecture is beyond the scope of this study which serves as a proof of principle; further theoretical and experimental work is clearly necessary.

\paragraph{Conclusions.} With the prediction of magnon NLSMs the correspondence of electronic and magnonic topologically nontrivial systems in ferromagnets is completed. The term `semimetal' does not respect the bosonic nature of magnons and, therefore, is strictly speaking meaningless. However, the topological features---here: nodal lines and drumhead surface states---exist irrespectively of fermion or boson statistics. The pendant of topological Dirac semimetals cannot be realized for magnons in ferromagnets because the twofold degeneracy of the bands is forbidden.

\begin{acknowledgments}
This work is supported by SPP 1666 of Deutsche Forschungsgemeinschaft (DFG).
\end{acknowledgments}

\bibliography{short,newrefs}
\bibliographystyle{apsrev}

\end{document}